\documentclass{PoS}

\newcommand{\bd}{\begin{displaymath}}
\newcommand{\ed}{\end{displaymath}}
\newcommand{\be}{\begin{equation}}
\newcommand{\ee}{\end{equation}}
\newcommand{\bda}{\begin{eqnarray*}} 
\newcommand{\eda}{\end{eqnarray*}}
\newcommand{\bea}{\begin{eqnarray}} 
\newcommand{\eea}{\end{eqnarray}}
\newcommand{\bi}{\begin{itemize}}
\newcommand{\ei}{\end{itemize}}
\def\slash#1{\mkern-1.5mu\raise0.4pt\hbox{$\not$}\mkern1.2mu #1\mkern 0.7mu}

\title{Chiral violations from one-loop domain wall fermions}

\ShortTitle{Chiral violations from one-loop domain wall fermions}

\author{
Stefano Capitani
\thanks{Address after October 1st, 2007: Institut f\"ur Kernphysik, 
        Universit\"at Mainz, Germany}\\
        Fakult\"at f\"ur Physik, Universit\"at Bielefeld, Germany\\
        E-mail: \email{capitani@physik.uni-bielefeld.de}}

\abstract{We present results from lattice perturbation theory for the 
residual mass and other matrix elements measuring the breaking of chiral 
symmetry in domain-wall fermions.

We have used the exact propagators corresponding to a finite number 
of points in the fifth dimensions, and results were obtained for 
several choices of the domain-wall parameters.}

\FullConference{The XXV International Symposium on Lattice Field Theory\\
                 July 30-4 August 2007\\
                 Regensburg, Germany}

\begin{document}

\section{Introduction}

Domain-wall simulations use lattices with a finite number of points $N_s$ in
the 5th dimension \cite{Boyle}, and so a breaking of chiral symmetry occurs.
Only in the theoretical limit in which $N_s=\infty$ the chiral modes can
fully decouple from each other, yielding an exact chiral symmetry.
Here we study these chiral violations using perturbative calculations and
computing three quantities: the residual mass $m_{res}$,
the difference $\Delta = Z_V - Z_A$, and $c_{mix}$, a chirally-forbidden
mixing (which is then nonzero at finite $N_s$) of an operator 
which measures the lowest moment of the $g_2$ structure function.

We have studied the dependence of these three quantities on $N_s$ and the 
domain-wall height $M$, and calculated the deviations from the $N_s=\infty$ 
results when $N_s$ is limited to small values, of $O(10)$.
We have hence repeated the computations for several choices of $N_s$ 
and $M$. A thorough exploration of large regions in the two-dimensional 
space spanned by $N_s$ and $M$ would be instead quite expensive for 
Monte Carlo simulations, and perturbation theory seems the more 
practical and cheaper way to gather hints of what is happening when 
the parameters are moved in this space.

In order to carry out these calculations one must use the Feynman rules 
which correspond to the theory truncated at finite $N_s$, and thus we 
also had to compute the required propagator functions. 

We have calculated the same quantities with the plaquette action 
\cite{Capitani:2006kw} as well as with improved gauge actions 
\cite{Capitani:2006yv}, since in numerical simulations it was
observed that these improved gauge actions (especially DBW2) reduce 
the chiral violations.
We refer to \cite{Capitani:2006kw,Capitani:2006yv} for the actions,
notations and conventions used, and in particular for the expressions of the
domain-wall fermion propagators at finite $N_s$. Here we only remind that 
$0<M<2$ and that the chiral projectors are $P_\pm = (1 \pm \gamma_5)/2$.

This domain-wall formulation \cite{Shamir:1993zy} corresponds to having 
several flavors of lattice Dirac fermions which are mixed via a mass matrix 
in a very special way, so that a large mass hierarchy is generated. 
To determine the chiral modes one must diagonalize (in the fifth dimension) 
this mass matrix, which however is not hermitian. Its square must then be
considered, which means the second-order operators $D D^\dagger$ and 
$D^\dagger D$. 
They are hermitian and nonnegative and give a well-behaved spectrum.

A rotation of the 5-dimensional quark fields $\psi_s(x)$ to the basis which 
diagonalizes the mass matrix gives finally the expression of the chiral mode:
\bd
\chi_0 (x) = \sqrt{1-w_0^2} \, \sum_s (P_+ w_0^{s-1} \psi_s (x) 
                               + P_- w_0^{N_s-s} \psi_s (x)) ,
\ed
where from now on we put $w_0 = 1 - M$. We can see from the damping factors 
$w_0^{s-1}$ and $w_0^{N_s-s}$ that the chiral mode is exponentially 
localized near the two walls at $s=1$ and $s=N_s$. However, the 
domain-wall height $M$, which is not protected by chiral symmetry, 
undergoes an additive renormalization, so that $w_0$ 
is also additively renormalized.

The standard chiral mode used in Monte Carlo simulations is then not 
$\chi_0 (x)$. It contains instead only the quark fields exactly located 
at the boundaries:
\bd
q(x) = P_+ \psi_1 (x) + P_- \psi_{N_s} (x), \qquad
\overline q(x) = \overline\psi_{N_s} (x) P_+ 
+ \overline\psi_1 (x) P_- .
\ed
These physical quark fields $q(x)$ are more convenient to use than 
$\chi_0 (x)$: they do not contain $w_0$ and avoid the problem 
of its renormalization.

At finite $N_s$ an additional issue arises: $\chi_0(x)$ itself is no longer 
the exact expression of the chiral mode. In fact, $\chi_0(x)$ at finite 
$N_s$ is an eigenvector of the mass matrix only up to terms of order 
$N_s \, e^{-N_s\alpha(0)}$, where $\alpha(0)$ is a constant determined by
$\,\,\,2\cosh (\alpha(0)) = (1+w_0^2)/|w_0|$.

\section{Residual mass at tree level}

The calculation of the propagator of the (approximate) chiral fields $q(x)$ 
gives (for $m=0$)
\bd
\langle q (-p) \overline q (p)\rangle = \frac{i 
\gamma_\mu \sin p_\mu \, (1-e^{-2N_s\alpha(p)}) 
+ e^{-N_s\alpha(p)} \cdot 2 W(p) \sinh (\alpha (p))}{1 - W(p) \, 
  e^{\alpha(p)} - e^{-2N_s \alpha(p)} \big(1-W(p) \, e^{-\alpha(p)}\big)} ,
\ed
where $\,\,\,W(p) = 1 - M + 2 \sum_\lambda \sin^2 \frac{p_\lambda}{2}\,\,\,$
and
$\,\,\,2\cosh (\alpha(p)) = (1+W^2(p)+\sum_\lambda \sin^2 p_\lambda)/|W(p)|$.
In the limit of small momentum this 4-dimensional propagator becomes
\bd
\langle q (-p) \overline q (p)\rangle \Big|_{p \ll 1} = - (1-w_0^2) \,\, 
\frac{i\slash{p} + w_0^{N_s} (1-w_0^2)}{p^2 + w_0^{2N_s}(1-w_0^2)^2} .
\ed
We can thus see that, although in the bare Lagrangian all quark fields are 
massless, the truncation of domain-wall fermions at finite $N_s$ generates 
already at the tree level a nonvanishing residual mass of the physical fields:
$a \, m_{res}^{(0)} = - w_0^{N_s} (1-w_0^2) = - (1-M)^{N_s} \, M(2-M)$.
As expected, this tree-level residual mass vanishes when 
$N_s$ becomes infinite\footnote{Indeed, since $w_0 = e^{-\alpha(0)}$, 
it is easy to see that the terms 
which are proportional to $w_0^{N_s} = e^{-N_s\alpha(0)}$ rapidly approach 
zero when $N_s$ becomes large.}.
Its sign can be inferred from the general expression of a fermion 
propagator of mass $\mu$ for small momentum in Euclidean space:
$(-i\slash{p} + \mu)/(p^2 + \mu^2) = 1/(i\slash{p} + \mu)$.
Since we work with even $N_s$ (where the fermion determinant 
can be proven to be positive), $m_{res}^{(0)}$ is always a 
negative quantity. With our calculations we have thus reproduced, 
up to a sign, the result for $m_{res}^{(0)}$ found in
\cite{Shamir:1993zy,Vranas:1997da,Vranas:1997ib,Kikukawa:1997tf,Blum:1999xi},
where it was derived by considering the quadratic operator $D^\dagger D$, 
which could perhaps explain the sign discrepancy.

\section{Physical propagator at one loop}

At one loop we can write\footnote{Evidencing the damping factors of the
external legs and the loop integral $\Sigma_{st} (p)$, the structure 
of $\Sigma_q (p)$ is 
\bea
\Sigma_q (p) & = \sum_{s=1}^{N_s} \sum_{t=1}^{N_s} 
& \frac{1}{1-w_0^{2N_s}} \, \Big[ 
   \big(w_0^{N_s-s} - w_0^{2N_s} w_0^{-(N_s-s)}\big) P_+ 
 + \big(w_0^{s-1} - w_0^{2N_s} w_0^{-(s-1)}\big)  P_- \nonumber \\
&& -w_0 \, \frac{i\slash{p}-w_0^{N_s}(1-w_0^2)}{1-w_0^2} 
\, \Big( \big(w_0^{s-1} - w_0^{2(N_s-1)} w_0^{-(s-1)} \big) P_+ 
       + \big(w_0^{N_s-s} - w_0^{2(N_s-1)} w_0^{-(N_s-s)} \big) P_- 
 \Big) \Big] \nonumber \\
&& \cdot \Sigma_{st} (p) \cdot \frac{1}{1-w_0^{2N_s}} \, \Big[ 
   \big(w_0^{N_s-t} - w_0^{2N_s} w_0^{-(N_s-t)}\big) P_- 
 + \big(w_0^{t-1} - w_0^{2N_s} w_0^{-(t-1)}\big)  P_+ \label{eq:dampfact} \\
&& - w_0 \, \Big( \big(w_0^{t-1} - w_0^{2(N_s-1)} w_0^{-(t-1)} \big) P_- 
       + \big(w_0^{N_s-t} - w_0^{2(N_s-1)} w_0^{-(N_s-t)} \big) P_+ 
 \Big) \, \frac{i\slash{p}-w_0^{N_s}(1-w_0^2)}{1-w_0^2} 
\Big] . \nonumber
\eea
}
\bda
\langle q (-p) \overline q (p)\rangle_{1~loop} 
& =& 
\frac{1-w_0^2}{i\slash{p}-w_0^{N_s}(1-w_0^2)} +
\frac{1-w_0^2}{i\slash{p}-w_0^{N_s}(1-w_0^2)} \, 
\Sigma_q (p) \, \frac{1-w_0^2}{i\slash{p}-w_0^{N_s}(1-w_0^2)}  
\nonumber \\
& = &  
\frac{1-w_0^2}{i\slash{p}-w_0^{N_s}(1-w_0^2)
- (1-w_0^2) \, \Sigma_q (p) } .
\eda

The general form of $\Sigma_q (p)$ for $m=0$ is (where we call for brevity 
$\bar g^2 = (g_0^2/16 \pi^2)\, C_F$)
\bd
\Sigma_q (p) =  \frac{\bar g^2}{1-w_0^2} \, \Big[ \frac{\Sigma_0}{a}  
+ i\slash{p} \, \Big( c_{\Sigma_1}^{(N_s,M)} \log a^2 p^2 + \Sigma_1 \Big)
- \big( i\slash{p}-w_0^{N_s}(1-w_0^2) \big) \, 
\frac{2w_0}{1-w_0^2} \, \Sigma_3 \Big] .
\ed
The most important difference in $\Sigma_q$ from its expression at infinite 
$N_s$ is the appearance of a totally new contribution, 
$\Sigma_0$, proportional to $1/a$ and associated with the breaking 
of chiral symmetry. $\Sigma_0$ comes from the terms of $\Sigma_{st}(p)$ 
which are of order zero in $p$, and acts as a mass correction term. Since
\bd
\langle q (-p) \overline q (p)\rangle_{1~loop} =
\frac{1-w_0^2}{i\slash{p} -w_0^{N_s}(1-w_0^2) -(1-w_0^2) \, 
\Sigma_q (p)} =
\frac{1-w_0^2}{i\slash{p} \, Z_2^{-1} + m_{res}^{(1)}} \, Z_w ,
\ed
the one-loop radiatively induced mass is given by
\bd
a \, m_{res}^{(1)} = - w_0^{N_s}(1-w_0^2) - \bar g^2 \, \Sigma_0 .
\ed
Thus, $\Sigma_0$ generates a finite additive renormalization to the 
residual mass when $N_s$ is not infinite\footnote{Of course higher loops 
and nonperturbative effects give further contributions to the shift 
of the residual mass.}. The factor 
$Z_w =  1 - 2 \bar g^2 \, \Sigma_3 \, w_0 / (1-w_0^2) =  1 + \bar g^2 \, z_w$
generates the additive renormalization to $w_0$ at this order 
\cite{Aoki:1998vv}, as we can see from
$(1-w_0^2) \, Z_w = 1 - \big( w_0 + \bar g^2 \, \Sigma_3 \big)^2 + O(\bar g^4)$.

The renormalization of a composite operator $\overline q(x) \, O \, q(x)$ 
which is multiplicatively renormalizable can also be expressed in a simple 
way:
\bd
\langle \, ( \, \overline q O q \, ) \, q \overline q \, \rangle_{1~loop} 
=\frac{1-w_0^2}{i\slash{p}-w_0^{N_s}(1-w_0^2)} \cdot A_O (p) \cdot 
\frac{1-w_0^2}{i\slash{p}-w_0^{N_s}(1-w_0^2)} ,
\ed
where $A_O (p)$ contains also the contribution of the 
damping factors, and takes the form
\bd
A_O (p) = \bar g^2 \Big( - \gamma_O^{(N_s,M)} \log a^2 p^2  + B_O \Big)
\ed
for a logarithmically divergent operator. The coefficients of the divergences
turn out to be different from their continuum values, and they depend on 
$N_s$ and $M$. It is only when $N_s=\infty$ that the anomalous 
dimensions become the ones calculated in the continuum theory.
In particular, the vector and axial-vector currents acquire a nonzero 
anomalous dimension at any finite $N_s$:
\bd
\gamma_V^{(N_s,M)} = 2 \, w_0^{2N_s} \, 
\Bigg( 1 - N_s \, w_0^{2N_s} \, \frac{1-w_0^2}{1-w_0^{2N_s}} \Bigg)
\, \Bigg( N_s  \, \frac{1-w_0^2}{1-w_0^{2N_s}} \, 
                 \Big( 2 + \frac{1}{1-w_0^{2N_s}} \Big)
     - 2 - \frac{w_0^2}{1-w_0^{2N_s}} \Bigg).
\ed
Furthermore, the residual mass as well as renormalization factors and mixing 
coefficients turn out to lose gauge invariance when $N_s$ is not infinite.
Although numerically the deviations from gauge invariance remain 
in most practical cases rather small, this is another of the
pathological features of the domain-wall theory truncated at finite $N_s$.
It could be that this is a limitation of perturbation theory, but it could be 
that a small gauge dependence is also present in numerical simulations.

These pathologies could actually be related to the mismatch between 
the (simplified) chiral modes which are actually used, and the true
chiral modes (the ones which contain $w_0$).
Notice that at finite $N_s$ there is an additional mismatch, because 
terms of order $N_s \, e^{-N_s\alpha(0)}$ and higher, which are present 
in the true chiral modes for $N_s<\infty$, are here missing as well. 
Thus, if calculations with the true chiral modes would be gauge invariant
and reproduce the continuum anomalous dimensions, the missing pieces 
from these mismatches could then account for the above pathologies.

At one loop two diagrams contribute to $\Sigma_0$ and so enter in the 
calculation of the residual mass: the half-circle (or sunset) and the 
tadpole diagrams. We have automated the calculations of the half-circle 
diagram (as well as the vertex diagrams for $\Delta$ and $c_{mix}$) by 
developing suitable FORM codes \cite{Vermaseren:2000nd}, integrating 
afterwards the corresponding expressions by means of Fortran codes.
With these programs we are able to compute matrix elements for general 
values of $N_s$ and $M$.

\section{The tadpoles}

The behavior of the tadpole diagrams as $N_s$ and $M$ change is 
particularly important.

The tadpoles do not contain pure 5-dimensional quark propagators, and so
for them the $\Sigma_{st} (p)$ of Eq. (\ref{eq:dampfact}) is diagonal 
in the fifth-dimensional index, and also proportional to 
$(i\slash{k} - 4r/a) \, G_{\mu\nu}(k)$, where $G_{\mu\nu}$ is the gluon 
propagator. This is the same integrand of the tadpoles for Wilson fermions, 
where in the case of the tadpole diagram contributing to $\Sigma_1$ 
it gives the result (in a general covariant gauge) 
$T_l = 8\pi^2 Z_0 \, (1-1/4\, (1-\lambda))$, with 
$Z_0=0.154933390231\ldots$ a well-known integral \cite{Capitani:2002mp}.

It is then clear that for domain-wall fermions the behavior of the tadpole 
diagrams as a function of $N_s$ (and $M$) is completely determined by 
the damping factors in the fifth dimension, (see Eq. (\ref{eq:dampfact})).
Their general effect can be already seen by looking at their 
leading contributions for large $N_s$. In this approximation
the damping factors enter the game in either of these combinations:
\bd
\sum_{s=1}^{N_s} w_0^{s-1} w_0^{N_s-s} = N_s w_0^{N_s-1}, \qquad
\sum_{s=1}^{N_s} (w_0^2)^{s-1} = \sum_{s=1}^{N_s} (w_0^2)^{N_s-s} = 
\frac{1-w_0^{2N_s}}{1-w_0^2} .
\ed
These are indeed the leading expressions, in units of
$T_d = (1-w_0^2) T_l / (1-w_0^{2N_s})^2$, 
for the tadpole contributions to (respectively) $\Sigma_0$ and $\Sigma_1$, 
in the limit of large $N_s$.
Already from these asymptotic expressions (before computing the exact 
results) we can immediately see that the tadpole of $\Sigma_0$ vanishes 
when $N_s=\infty$, while the tadpole of $\Sigma_1$ 
gives in this limit the known Wilson number, $T_l$.

Thus, the damping factors play a primary r\^ole in determining the values
of the domain-wall tadpoles. After calculating their exact expressions, 
which include all subleading terms in $N_s$, the tadpole 
contribution to $\Sigma_0$ turns out to be equal to
\bd
  4 \, T_d \, \Big[ N_s \, (1 + w_0^{2(N_s+1)}) \, w_0^{N_s-1} 
       - 2 \, w_0^{N_s+1} \, \frac{1-w_0^{2N_s}}{1-w_0^2} \Big] ,
\ed
while the tadpole contribution to $\Sigma_1$ turns out to be equal to
\bd
T_d \, \Big[ (1 + w_0^{2(N_s+1)}) \, \frac{1-w_0^{2N_s}}{1-w_0^2} 
       - 2 \, N_s \, w_0^{2N_s} \Big] .
\ed
The values of these tadpoles present wide variations with $N_s$ and $M$, 
so that sometimes they turn out to be small while in other situations 
they become large.
This suggests that some care should be used when talking about tadpole 
dominance in relation to domain-wall fermions. It also happens that
both tadpoles (of $\Sigma_0$ and $\Sigma_1$) even decrease toward zero 
for $M\to 0$ or $M\to 2$.

A central point is that there are two kinds of tadpoles in the game here:
\bi
\item the tadpole of order zero in $p$, which 
tends to zero for $N_s \to \infty$,  
and which contributes to $\Sigma_0$ and the residual mass;
\item the tadpole of order $ap$, which 
tends to its Wilson value for $N_s \to \infty$, 
and which contributes to $\Sigma_1$ and the renormalization factors.
\ei
They behave quite differently, and this is because the $i\slash{p}$ of the 
first order of the self-energy flips the chirality of some damping factors, 
which then combine in a different way.
We stress that this is quite unlike the Wilson case, where the tadpole of 
$\Sigma_0$ is just proportional to the tadpole of $\Sigma_1$:
\bd
T_{(\Sigma_0)}^{(Wilson)} = -4 \, T_{(\Sigma_1)}^{(Wilson)} .
\ed

Tadpole improvement seems then not to be appropriate for the residual mass: 
the tadpole which contributes to $m_{res}$ goes to zero for large $N_s$ or 
for $M \to 1$, and for small $N_s$ it assumes a wide spectrum of values.
Our interest is small $N_s$, where it is unclear what tadpole improvement 
(or resummations) could mean. Moreover, no tadpole enters at all in the 
calculations of $\Delta$ and $c_{mix}$.

For large $N_s$ the tadpole of $\Sigma_1$ is rather close to its Wilson 
value, and that is why tadpole improvement could be used in the 
calculations of the renormalization factors (in the large $N_s$ limit).

\section{Residual mass at one loop}

Our one-loop perturbative calculations show that the numerical deviations 
from the case of infinite $N_s$ depend, apart from $N_s$ (and to a smaller 
extent from $g_0$), very strongly on the choice of $M$.
We can observe that the deviations from the case of exact chiral symmetry 
are rather pronounced when $M \sim 0.1$ or $M \sim 1.9$. The values of 
$a m_{res}^{(1)}$ turn out to be positive only for $M \ge 1.2$ (at least 
for even $N_s$ and if the coupling is not very small), and 
our results suggest that the minimal amount of chiral violations 
is attained for $M \sim 1.2$. This is then the optimal choice of $M$ 
from the point of view of one-loop calculations, and corresponds
to the renormalization of $M$, which is not protected by 
chiral symmetry and is then moved by radiative corrections away from its 
free field value $M=1$.
One can conjecture that higher-loop corrections and nonperturbative effects 
would shift this optimal value further on, until the minimal point 
is eventually reached around $M \sim 1.8$ (which provides the smallest 
residual mass in Monte Carlo simulations).

We also can observe that for $M=1.9$ the residual mass at $N_s=12$ is larger 
than at $N_s=8$, and at $N_s=16$ is even larger. For a detailed discussion 
of these phenomena which occur near the borders of the allowed values for $M$
we refer to \cite{Capitani:2006kw}.

With improved gauge actions we can still see that the residual mass 
$a m_{res}^{(1)}$ is positive only for $M \ge 1.2$.
This also shows that improved gauge actions do not behave too differently 
in terms of the additive renormalization undergone by $w_0$. 
Employing improved gauge actions produces, not surprisingly, a suppression 
of $m_{res}$ when one carries out the comparisons at the same value of the 
coupling. The Iwasaki action gives a stronger suppression than the 
L\"uscher-Weisz action, and there seems to be a monotonic decrease 
of the residual mass as $c_1$ grows. The DBW2 action is indeed 
the most effective in generating large suppressions.

If comparisons between the various actions are instead made at the same 
energy scale, the picture that comes out is different from naive expectations.
For example, for quenched QCD at 2 GeV one has to take $\beta=5.7$ for the 
L\"uscher-Weisz action, $\beta=2.6$ for the Iwasaki action,
and $\beta=1.04$ for the DBW2 action\footnote{We use $\beta = 6/g_0^2$ 
also for improved actions, instead of $\beta' = 6 (1-8 c_1)/g_0^2$.}.
The $m_{res}$ numbers for the Iwasaki action are then rather close 
to those of the DBW2 action, and surprisingly they lie in general slightly 
above the plaquette values.

However, for the quenched DBW2 action at 2 GeV one has $g_0^2 = 5.77$,
which is rather large, and so the one loop results cannot perhaps 
be trusted so easily\footnote{In this case the L\"uscher-Weisz action 
gives the largest $m_{res}$ suppression, and indeed $g_0^2$ at 2 GeV 
is still close to $1$.}.
Moreover, these values of the couplings are determined from numerical 
simulations, and they then contain informations of a nonperturbative 
nature, so that a mismatch can arise when one only takes into account 
the results of the one-loop diagrams calculated for these values of the 
couplings.

Many numerical results, which we cannot include here for lack of space, 
can be found in \cite{Capitani:2006kw,Capitani:2006yv}.

\section{Bilinear differences, and a power-divergent mixing}

Since $Z_V \neq Z_A$ when chiral symmetry is broken, the difference between 
these renormalization constants, $\Delta = Z_V - Z_A = -(Z_S - Z_P)/2$, 
provides an estimate of chirality-breaking effects.

The amount of chirality breaking connected to $\Delta$ follows a 
pattern similar to the one of the residual mass: $\Delta$ is rather large 
for small $N_s$ or $|1-M|\sim 1$, it decreases when $N_s$ grows or when 
$|1-M|$ tends towards zero, and the violations of gauge invariance
are very small.

The numbers for $\Delta$ come out much smaller, at a given $M$ and $N_s$, 
than the ones for the residual mass. For quantities such as four-quark 
operators it was suggested in \cite{Aoki:2004ht,Christ:2005xh}
that their chiral violations are of $O(m_{res}^2)$.
Given the smallness of the numbers that we have obtained for $\Delta$, 
it is possible that something similar is also occurring here.

We have also calculated the mixing of the antisymmetric operator 
\bd
O_{d_1} = \bar q (x) \, \gamma_{[4} \gamma_5 D_{1]} \, q (x)
\ed
with an operator of lower dimension, 
\bd
c_{mix} \cdot \frac{i}{a} \, \bar q (x) \, \sigma_{41} \gamma_5 \, q (x).
\ed
The operator $O_{d_1}$ enters in the calculation of the first moment of 
the $g_2$ structure function, and has been simulated using quenched 
domain-wall fermions with the DBW2 gauge action \cite{Orginos:2005uy}.

The power-divergent mixing of $O_{d_1}$ on the lattice is only caused by 
the breaking of chirality, and hence it provides a quantitative measure 
of chiral violations. In the theoretical limit $N_s=\infty$ one has instead $
c_{mix}=0$ and so $O_{d_1}$ becomes multiplicatively renormalized.

The chiral violations associated with $c_{mix}$ are rather small,
and thus they also seem to be of higher order in $m_{res}$. The pattern of 
the deviations from the case of exact chirality is the usual one.


\begin{thebibliography}{99}
\bibitem{Boyle}
  For an overview of recent Monte Carlo results with domain-wall fermions 
  we refer to the plenary talk of Peter Boyle in these Proceedings, 
  and to the references therein.
\end{thebibliography}
\end{document}